# The influence of the degree of deformation-induced martensitic phase and structural transformation fcc to bcc on the low-temperature dynamic Young's modulus of the medium-entropy alloy $Co_{17.5}Cr_{12.5}Fe_{55}Ni_{10}Mo_5$

Y. Semerenko [a], E. Tabachnikova [a], T. Hryhorova [a], S. Shumilin [a], Y. Shapovalov [a], H. Kim [b], J. Moon [b], H. Kwon [b], and V. Zoryansky [a]

[a] *B. Verkin Institute for Low Temperature Physics and Engineering of the NAS of Ukraine, Kharkiv, Ukraine*
[b] *Pohang University of Science and Technology (POSTECH), 77 Cheongam-Ro, Nam-gu, KR-37673 Pohang, Gyeongbuk, Republic of Korea*

**Abstract.** The elastic properties and microstructural evolution of a medium-entropy alloy $Co_{17.5}Cr_{12.5}Fe_{55}Ni_{10}Mo_5$ (at%) in a low temperature range were investigated. It has been established that low-temperature plastic deformation initiates martensitic phase transformations in this alloy, and the values of the dynamic modulus of elasticity correlate with the degree of phase transformations.

**Introduction**

Over the past two decades, a new class of high-entropy alloys (HEAs) have been designed [1]. Investigation of this materials has grown explosively [1–7]. HEAs - solid-state metal systems of five or more components with a concentration close to equiatomic. Such alloys are characterized by increased values of the entropy of mixing *S*, compared to traditional multicomponent alloys, which explains their name. The explanation for the increased entropy value in HEAs is based on the concepts of classical thermodynamics that the entropy of mixing between soluble components is maximum when these components are in equiatomic concentration and increases with increasing number of components [8]. The change in Gibbs free energy $\Delta G$ when mixing HEA components is determined by the relation:

$$\Delta G = \Delta H - T\Delta S, \tag{1}$$

where ΔH is the change in enthalpy of the system, *T* is the temperature.

Thus, the significant contribution of the entropy of mixing during the formation of HEAs reduces the free energy, as a result of which the probability of the formation of substitutional solid solutions with simple crystal lattices in them significantly increases. Indeed, many HEAs have the structure of single-phase solid solutions with face-centered cubic (**fcc**) or body-centered cubic (**bcc**) crystal lattices, and such lattices are significantly distorted, since they are formed by atoms of dissimilar elements with different electronic structures and sizes

Among these materials, **fcc** HEAs stand out as they exhibit excellent mechanical properties at cryogenic temperatures owing to deformation-induced twinning [2, 9] and/or martensitic phase transformations [3, 4, 10]. Specifically, the 'metastability engineering' approach to martensitic transformations, whereby phase stability is controlled through chemical composition and deformation temperature, has been widely used [3, 4, 11]. To strengthen HEAs at cryogenic temperatures, Bae *et al.* [4] designed metastable ferrous HEAs ($Fe_x(CoNi)_{90-x}Cr_{10}$, x = 55-60 at%) with excellent strain hardening and tensile strength characteristics at low temperature owing to deformation-induced martensitic transformation from **fcc** to **bcc** crystal structure. However, the yield strength of ferrous HEAs is relatively low because the initial microstructures of the alloys contain an fcc single phase [4]. The addition of molybdenum to ferrous HEAs enhances yield strength due to the precipitation strengthening by µ phase in the **fcc** matrix [6]. Therefore, the strategy of "*metastability engineering*" as applied to Mo-added ferrous HEAs appears very hopeful. It is expected to provide a favorable combination of high strain hardening with excellent yield strength and tensile strength at cryogenic temperatures.

Recent studies have shown that a some of non-equiatomic metastable high-entropy (HEA) and medium-entropy alloys (MEA) have a promising ratio of strength and ductility, especially at cryogenic temperatures [7]. Recently, a metastable nonequiatomic MEA $Co_{17.5}Cr_{12.5}Fe_{55}Ni_{10}Mo_5$ was developed on the basis of iron and molybdenum additives (the indices correspond to the atomic concentration). In this MEA, a favorable combination of yield strength and ultimate strength is observed due to the action of phase martensitic transformations initiated by plastic deformation (DIMT) [6]. Previous works [6, 9, 10] have studied the structure of this MEA in detail, as well as the effect of plastic deformation at different temperatures on DIMT. It was found [9] that tensile-deformed $Co_{17.5}Cr_{12.5}Fe_{55}Ni_{10}Mo_5$ alloy, due to DIMT from fcc to bcc structure, has excellent low-temperature mechanical properties. So, for example, at 4.2 K the yield strength is 1043 MPa, and the tensile strength is 1748 MPa, while maintaining high plasticity.

It is known that the elastic characteristics of a material are highly sensitive to various phase and structural changes. Therefore the aim of this work was to study the effect of DIMT on the elastic characteristics of the molybdenum doped alloy.

**Research methods, sample characteristics**

In a wide range of temperatures, the acoustic and structural properties of a medium-entropy alloy with the nominal composition of components $Co_{17.5}Cr_{12.5}Fe_{55}Ni_{10}Mo_5$ were investigated. Liquid $^4$He (T=4.2 K) and liquid nitrogen (T=77 K) were used to obtain cryogenic temperatures. Intermediate temperatures in the range of 4.2–77 K were obtained by cooling the samples with helium vapor.

Samples of $Co_{17.5}Cr_{12.5}Fe_{55}Ni_{10}Mo_5$ alloy was obtained by vacuum induction melting in an Ar atmosphere with metals with a purity of over 99.9% by the standard procedure described in [8]. To reduce the grain, the ingots were rolled at room temperature until the thickness was reduced by 79%

(from 7 to 1.5 mm). The rolled sheets were annealed at a temperature of 900° C for 60 minutes, and then quenched in water.

Samples for further research were cut by electroerosion cutting from larger blanks in the direction of rolling, and then mechanically ground and polished with abrasive powders until the required shape and dimensions were achieved. After that, the alloy $Co_{17.5}Cr_{12.5}Fe_{55}Ni_{10}Mo_5$ samples were mechanically deformed. Plastic deformation of dog-bone shaped tensile samples with a gauge length of 15 mm and a width of 3 mm was carried out at temperatures of 77 K, 4.2 K, 2.1 K, and 0.5 K by uniaxial tension to fracture (a true strain was about of 30%) at a constant rate $10^{-4}$ s$^{-1}$. Next, samples for acoustic and structural studies were cut from the deformed samples. The test samples for acoustic investigations were cut by electric spark and reduced to the final size of 0.3×4.4×20.3 mm$^3$ by mechanical polishing. After the elastic properties of the undeformed sample were studied, this sample was deformed by bending at a temperature of 77 K (a true strain was about of 10%) and its elastic properties were measured again.

The temperature dependences of the dynamic Young modulus $E$ were measured at temperatures of 80-280 K. The rate of change of the temperature was ≈1 K/min and the accuracy of the temperature measurements was ≈50 mK.

The elastic characteristics of the alloy were investigated in the temperature range of 80-280 K by mechanical resonance spectroscopy [11].

Acoustic measurements were performed using mechanical resonance spectroscopy. Linear bending vibrations of a cantilevered sample were studied at small values of the deformation amplitude $\varepsilon_0 \sim 10^{-7}$. Under the influence of a cyclic electric force, forced vibrations of the thin plate were excited with a variation of its frequency $f$ near the first resonant frequency of mechanical vibrations $f_r = 530$ Hz in the amplitude independent deformation region. As a result, an alternating stress $\sigma = \sigma_0 \sin(\omega t)$ acted on the sample of material with a cyclic frequency $\omega = 2\pi f$, and the sample under study experienced a small cyclic deformation $\varepsilon_0 \cdot \sin(\omega t - \varphi)$, which, due to the strain viscosity of the sample material, lagged in phase from the applied stress by an angle $\varphi$. In this case, $tg\varphi$ characterizes the absorption of the energy of mechanical vibrations in the studied material, and the dynamic modulus of elasticity is determined by the deformation component that is in phase with the applied stress. Oscillations of sample were detected electrostatically. A resonance occurs as the frequency $f$ of the external electrostatic driver force approaches the characteristic frequency $f_r$ of the mechanical oscillations of the sample. Then the elastic modulus $E$ of the cantilever mounted sample of given thickness $h$ and length $l$ depends on the experimentally measured resonance frequency $f_r$ of the mechanical oscillations of the test sample and is given by [11-13]:

$$E = 38.3 \frac{f_r^2 \rho l^4}{h^2}, \tag{2}$$

where $\rho \simeq 8.2$ g/cm$^3$ is the density of the sample and 38.3118 is a correction factor that depends on the shape of the sample and the Poisson coefficient (taken to be $\approx 0.3$).

**Results and discussion**

It was shown [6] that the structure of the annealed Co$_{17.5}$Cr$_{12.5}$Fe$_{55}$Ni$_{10}$Mo$_5$ alloy in the undeformed state is fully recrystallized and randomly oriented. The inverse pole figure (IPF) map shows that the annealed alloy is fully recrystallized and randomly oriented. The average grain size of the annealed alloy is 3.82 ± 1.82 µm. The X-ray diffraction (XRD) pattern reveals that the annealed alloy comprises an **fcc** matrix with submicron µ precipitates. The backscattered electron (BSE) image also shows that fine µ precipitates are dispersed in the matrix. The average size and the area fraction of the µ precipitates are 177.16±62.53 nm and 3.21±0.26%, respectively [8]. These µ-precipitates are identified as a polycomponent intermetallic phase with an **hcp** structure that contains all components of the alloy.

The phase composition of both undeformed and deformed samples was investigated. Phase images of electron backscatter diffraction (EBSD) for alloy deformed at cryogenic temperatures show that DIMT from **fcc** to hexagonal close packed (**hcp**) and **bcc** phase occurs. At true deformation of 10%, the **fcc** phase begins to transform in the **hcp** and **bcc** phase (see Fig. 1). With further deformation, DIMT continues and the proportion of **bcc** phase increases significantly, while the proportion of **hcp** phase remains below 10%. The deformation-induced martensitic transition of the **fcc** phase to **bcc** is almost complete with a true deformation of 30% at a temperature of 77 K, when the proportion of **bcc** phase is 98.2% (Table 1). Interestingly, the proportion of **bcc** at 30% strain decreased to 68.8% when the alloy deformed at 4.2 K, but increased to 80.3% and 87.7% when the strain temperature was reduced to 2.1 K and 0.5 K (see Table 1). The course of the temperature dependence of the dynamic Young's modulus of both undeformed and deformed samples coincides with the notions of the additive contribution of the phonon and electron components. An increase in temperature leads to a monotonic decrease in the dynamic modulus of elasticity by 8%.

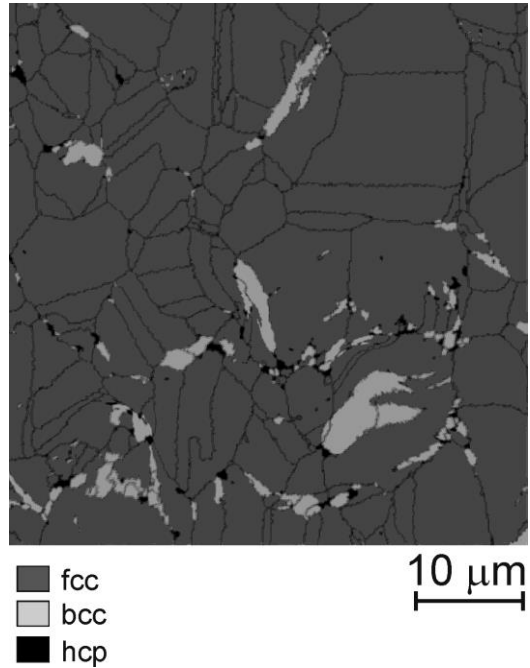

Fig. 1 EBSD phase map (Tescan Mira 3 LMU); bending deformation 10% at a temperature of 77 K.

Table 1. Dependence of the conversion rate of partial **fcc** – **bcc** DIMT on the deformation conditions.

| Part of the phase, % | Deformation mode and temperature of the deformation, K | | | | | |
|---|---|---|---|---|---|---|
| | Undeformed [9, 10] | bending, true strain 10% | uniaxial tension to fracture, true strain 30% [9, 10] | | | |
| | | 77.0 K | 4.2 K | 2.1 K | 0.5 K | 77.0 K |
| **fcc** | 100.0 | 89.8 | 30.7 | 13.4 | 8.8 | 1.3 |
| **bcc** | 0.0 | 5.0 | 68.8 | 80.3 | 87.7 | 98.2 |
| **hcp** | 0.0 | 5.2 | 0.5 | 6.3 | 3.5 | 0.5 |

Thus, it was established that undeformed alloy samples are single-phase and consist of 100% **fcc** phase; significant deformation under the most favorable conditions (strain to failure at 77 K) leads to an almost complete structural-phase transformation **fcc → bcc** of the entire volume of the sample; and less favorable deformation conditions lead to a partial **fcc → bcc** structural-phase transformation of the sample material. In this case, randomly oriented grains of different phases are uniformly distributed throughout the entire volume of the sample. This gives grounds to consider such a material as a multicomponent composite. Then the effective modulus of elasticity of such a material will be determined by averaging the modulus of elasticity of its components [14]. Among the existing models of such averaging, the approaches of Voigt and Reiss are the most productive, simple and universal. They make it possible to estimate, respectively, the upper and lower limits of the effective values of the elastic modulus of the composite based only on an analysis of the ratio of the volume fractions of the components without considering the features of the microstructure of the composite material.

Voigt's approach is based on the assumption of uniform deformation of the composite material, then the upper limit of the elastic modulus will be determined by the phase mixing rule [15]:

$$M_V = \sum c_i M_i, \qquad (3)$$

where $M_i$, $c_i$ - respectively, the elastic modulus of the $i$-th component and its volume fraction.

The Reiss approach assumes stress homogeneity and gives an estimate for the lower limit of the elastic modulus [16]:

$$\frac{1}{M_R} = \sum \frac{c_i}{M_i} \qquad (4)$$

Taking into account the insignificant fraction of the **hcp** phase in all the samples we studied, we can, as a first approximation, neglect the influence of its contribution to the total elastic modulus of the material under study and limit ourselves to the approximation of a two-component composite, in this case:

$$M_F = c_{fcc} M_{fcc} + c_{bcc} M_{bcc} = c_{bcc}\left(M_{bcc} - M_{fcc}\right) + M_{fcc}, \qquad (5)$$

$$M_R = \frac{M_{fcc} M_{bcc}}{c_{bcc}\left(M_{fcc} - M_{bcc}\right) + M_{bcc}}; \qquad (6)$$

where $M_{fcc}$ and $M_{bcc}$ are the elastic moduli of the **fcc** and **bcc** phases, respectively, and $c_{fcc}$ and $c_{bcc} = 1 - c_{fcc}$ are the volume fractions of these phases.

The arithmetic average of $M_V$ and $M_R$ called the Voigt-Reuss-Hill modulus is taken as the effective elastic modulus of the material [17] and provides a simple way to estimate the elastic properties of a multiphase material:

$$M_{VRH} = \frac{M_V + M_R}{2} \qquad (7)$$

In Fig. 2 shows the experimental values of the elastic modulus of samples with different degrees of **fcc** → **bcc** transformation, as well as the theoretical value of the effective elastic modulus $M_{VRH}$ calculated based on the approach outlined above. In this case, the elastic modulus values measured in an undeformed sample were taken as the $M_{fcc}$; and the values of the elastic modulus of a sample deformed at a temperature of 77 K until failure were taken as the $M_{bcc}$.

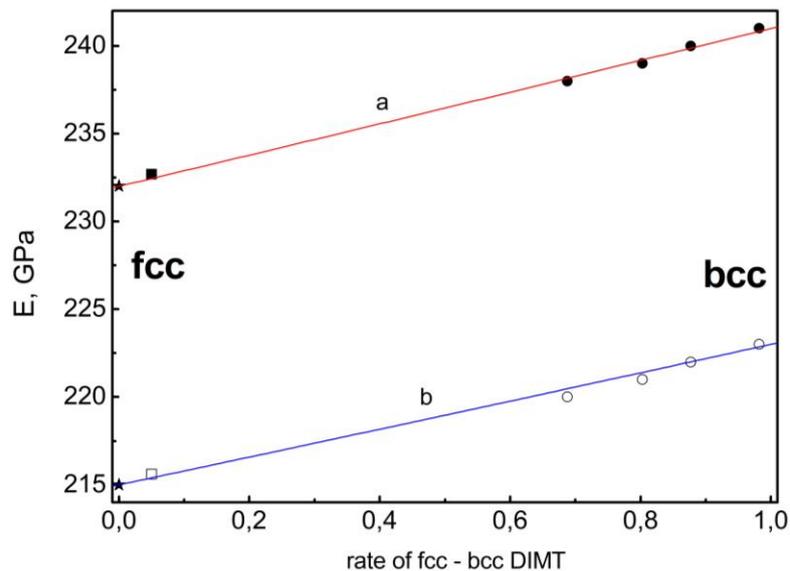

Fig. 2 Dependence of the dynamic Young's modulus $E$ of the $Co_{17.5}Cr_{12.5}Fe_{55}Ni_{10}Mo_5$ alloy on the rate of **fcc** – **bcc** DIMT (★ - undeformed sample [18]; ●, ○ - uniaxial tension to fracture, true strain 30% [18]; ■, □ - bending, true strain 10%): a) - dynamic Young's modulus $E$ at the 80 K; b) - dynamic Young's modulus $E$ at the 280 K. Solid lines – theoretical dependencies $M_{VRH}$ calculated using the formula (7).

**Summary**

The values of the dynamic modulus of elasticity in deformed samples of $Co_{17.5}Cr_{12.5}Fe_{55}Ni_{10}Mo_5$ alloy differ from the undeformed state and correlate with the degree of DIMT (fraction of bcc phase) – increases almost linearly as the proportion of the bcc phase increases (see Fig. 2). This dependence is described with good accuracy in the two-component composite approximation in accordance with the Voigt-Reiss-Hill model (7).

**References**

[1]  J.-W. Yeh, S.-K. Chen, S.-J. Lin, J.Y. Gan, T.-S. Chin, T.-T. Shun, C.-H. Tsau, S.-Y. Chang, Nanostructured high-entropy alloys with multiple principal elements: novel alloy design concepts and outcomes, Adv. Eng. Mater. **6** (2004) 299-303.

[2]  B. Gludovatz, A. Hohenwarter, D. Catoor, E.H. Chang, E.P. George, R.O. Ritchie, A fracture-resistant high-entropy alloy for cryogenic applications, Science **345** (2014) 1153-1158.

[3]  Z. Li, K.G. Pradeep, Y. Deng, D. Raabe, C.C. Tasan, Metastable high-entropy dual-phase alloys overcome the strength-ductility trade-off, Nature **534** (2016) 227-230.

[4]  J.W. Bae, J.B. Seol, J. Moon, S.S. Sohn, M.J. Jang, H.Y. Um, B.-J. Lee, H.S. Kim, Exceptional phase-transformation strengthening of ferrous medium-entropy alloys at cryogenic temperatures, Acta Mater. **161** (2018) 388-399.

[5]  X.D. Xu, P. Liu, Z. Tang, A. Hirata, S.X. Song, T.G. Nieh, P.K. Liaw, C.T. Liu, M.W. Chen, Transmission electron microscopy characterization of dislocation structure in a face-centered cubic high-entropy alloy $Al_{0.1}CoCrFeNi$, Acta Mater. **144** (2018) 107-115.

[6]     J.W. Bae, J.M. Park, J. Moon, W.M. Choi, B.-J. Lee, H.S. Kim, Effect of µ-precipitates on the microstructure and mechanical properties of non-equiatomic CoCrFeNiMo medium-entropy alloys, J. Alloys Compd. **781** (2019) 75-83.

[7]     J. Miao, C.E. Slone, T.M. Smith, C. Niu, H. Bei, M. Ghazisaeidi, G.M. Pharr, M.J. Mills, The evolution of the deformation substructure in a Ni-Co-Cr equiatomic solid solution alloy, Acta Materialia **132** (2017) 35-48.

[8]     D.A. Porter, K.E. Easterling, Phase transformations in metals and alloys. London: Chapman & Hall, 1992

[9]     J. Moon, E. Tabachnikova, S. Shumilin, T. Hryhorova, Y. Estrin, J. Brecht, P.K. Liaw, W. Wang, K.A. Dahmen, A. Zargaran, J.W. Bae, H.-S. Do, B.-J. Lee, H.S. Kim, Deformation behavior of a Co-Cr-Fe-Ni-Mo medium-entropy alloy at extremely low temperatures, Materials Today **50** (2021) 55-68.

[10]    Yu.O. Semerenko, E. D. Tabachnikova, T. V. Hryhorova, S. E. Shumilin, Yu.O. Shapovalov, H. S. Kim, J. Moon, H. Kwon, Low-Temperatures Physical-Mechanical Properties of the Medium-Entropy Alloy $Co_{17.5}Cr_{12.5}Fe_{55}Ni_{10}Mo_5$, Metallofiz. Noveishie Tekhnol. **43** (2021) 273-287 (in Ukrainian).

[11]    V.D. Natsik, Yu.A. Semerenko, Dislocation mechanisms of low-temperature acoustic relaxation in iron, Low Temp. Phys. **45** (2019) 551–567.

[12]    E.D. Tabachnikova, M.A. Laktionova, Yu.A. Semerenko, S.E. Shumilin, and A.V. Podolskiy, Mechanical properties of the high-entropy alloy $Al_{0.5}CoCrCuFeNi$ in various structural states at temperatures of 0.5–300 K, Low Temp. Phys. **43** (2017) 1108-1118.

[13]    Yu.A. Semerenko, Interfacing the Instrumental GPIB with a Personal Computer Through the LPT Port, Instr. Exp. Tech. **48** (2005) 608-610.

[14]    G. Grimvall, Thermophysical properties of materials (Elsevier Science B.V., Amsterdam, 1999).

[15]    W. Voigt, *Lehrbuch der Kristallphysik* (B. B. Teubner, Leipzig, 1928), p. 739.

[16]    A. Reuss, *Z. Angew. Math. Mech.* 9, 49 (1929).

[17]    CS. Man, M. Huang, A Simple Explicit Formula for the Voigt-Reuss-Hill Average of Elastic Polycrystals with Arbitrary Crystal and Texture Symmetries. *J Elast* **105**, 29–48 (2011). https://doi.org/10.1007/s10659-011-9312-y

[18]    Yu.O. Semerenko, E.D. Tabachnikova, T.V. Hryhorova, S.E. Shumilin, Yu.O. Shapovalov, H.S. Kim, J. Moon, H. Kwon, Influence of plastic deformation on the structure and dynamic module of elasticity of medium entropopy alloy $Co_{17.5}Cr_{12.5}Fe_{55}Ni_{10}Mo_5$ // VI International conference «High purity materials: Production, application, properties» Dedicated to memory of Academician of NASU V.M. Azhazha 13-15 September 2021, Kharkov, Ukraine